\documentstyle[12pt]{article}
\def\pmb#1{\setbox0=\hbox{$#1$}
\kern-.025em\copy0\kern-\wd0
\kern.05em\copy0\kern-\wd0
\kern-.025em\raise.0433em\box0 }
\def\bq{\begin{equation}}
\def\eq{\end{equation}}
\def\bqy{\begin{eqnarray}}
\def\eqy{\end{eqnarray}}
\def\bqyn{\begin{eqnarray*}}
\def\eqyn{\end{eqnarray*}}

\begin{document}

\begin{center}

{\Large {\bf Formation and Primary Heating of \\
The Solar Corona --- Theory and Simulation}}

\vspace{0.7cm}

{\bf S.M. Mahajan$^{\,a)}$} \\
\vspace{-0.2cm}
{\small \it Institute for Fusion Studies, The Univ. of Texas, Austin, 
TX
78712}\\

{\bf R. Miklaszewski$^{\,b)}$}\\
\vspace{-0.2cm}
{\small \it Institute of Plasma Physics and Laser Microfusion, 00--908
Warsaw, Str. Henry 23,\\
\vspace{-0.2cm}
P.O. Box 49, Poland }\\

{\bf K.I. Nikol'skaya$^{\,c)}$}\\
\vspace{-0.2cm}
{\small \it Institute of Terrestrial Magnetism, Ionosphere and Radio 
Wave
Propagation(IZMIRAN),\\
\vspace{-0.2cm}
Troitsk of Moscow Region, 142092 Russia}\\

{\bf N.L. Shatashvili$^{\,d)}$}\\
\vspace{-0.2cm}
{\small \it Plasma Physics Department, Tbilisi State University, 
380028,
Tbilisi, Georgia, \\
\vspace{-0.2cm}
International Centre for Theoretical Physics, Trieste, Italy}\\

\end{center}

\vspace{1cm}

{\small An integrated Magneto--Fluid model, that accords full treatment to the Velocity fields associated with the directed plasma motion, is developed to investigate the dynamics of coronal structures. It is suggested that the interaction of the fluid and the magnetic aspects of plasma may be a crucial element in creating so much diversity in the solar atmosphere. It is shown that the structures which comprise the solar corona can be created by particle (plasma) flows observed near the Sun's surface --- the primary heating of these structures is caused by the viscous dissipation of the flow kinetic energy.} 

\vfill

\begin{flushleft}
\rule{5.1 in}{.007 in} \\
$^{a)}$  \hskip 0.2cm {\small Electronic mail: \ 
mahajan@mail.utexas.edu}\\
\vspace{-0.3cm}
$^{b)}$  \hskip 0.2cm {\small Electronic mail: \ 
rysiek@ifpilm.waw.pl}\\
\vspace{-0.2cm}
$^{c)}$  \hskip 0.2cm {\small Electronic mail: \ 
hellab@izmiran.troitsk.ru}\\
\vspace{-0.2cm}
$^{d)}$  \hskip 0.2cm {\small Electronic mail: \  nanas@iberiapac.ge 
\hskip
0.3cm shatash@ictp.trieste.it}\\
\end{flushleft}

\newpage

\section{INTRODUCTION}

The TRACE observations [1--3] reveal that the solar corona is comprised of lots of thin loops that are intrinsically dynamic, and that continually evolve. These very thin strings, the observations indicate, are heated for a few to tens of 
minutes, after which the heating ceases, or at least changes significantly in 
magnitude [1]. In this  paper we  examine a class of mechanisms, which, through the viscous--dissipation of the plasma kinetic energy, provide the primary and basic heating of the coronal structures during their very formation. The basic input of the theory is the reasonable assumption that the coronal structures are created from the evolution and re--organization of a relatively cold plasma flow [1--16] emerging from the sub--coronal region (between the solar surface and the visible corona) and interacting with the ambient magnetic field anchored inside the solar surface. During the process of trapping and accumulation, a part of the kinetic energy of the flow is converted to heat by  viscous dissipation and the coronal structure is  born hot and bright. For this to happen, we must find alternative fast and efficient heating  mechanisms because, for the conditions prevalent in the coronal structures, the standard viscous dissipation is neither efficient nor fast. The rates of viscous dissipation can be considerably increased by processes which either enhance the local viscosity coefficient, or induce short scale structures in the velocity field. At present  we do not know  of any convincing mechanism for the former possibility. This paper, therefore, is limited to an examination of processes of the latter kind. We find that as long as the flow--velocity field is treated as an essential and integral part of the plasma dynamics, fast and desirable viscous dissipation does, indeed, result. Consequently, during its very formation, the  coronal structure can become hot and bright.

Of the several possible mechanisms by which the flow kinetic energy may be converted into heat we emphasize the following two:
The first is the ability of supersonic flows to create nonlinear perturbations which steepen to produce short scale structures which can dissipate by ordinary viscosity. The second stems from the recently established property of the magnetofluid equilibria for extreme sub--Alfv\'enic flows (most of the observed coronal flows fall in this category) -- such flows can have a substantial, fastly varying (spatially) velocity field component even when the magnetic field is mostly smooth. Viscous damping associated with this varying component could 
be a major part of the primary heating needed to create and maintain the bright Corona. From a general framework describing a plasma with flows, we have been able to ``derive'' several of the essential characteristics of the coronal structures. Theoretical basis for both these mechanisms will be discussed. Our simulation (for which we developed a dissipative two--fluid code), however,  concentrates only on the first mechanism, and preliminary results reproduce many of the salient observational features. There is clear cut evidence of nonlinearly steepened velocity fields which effectively dissipate and heat the coronal structure right through the process of formation. The numerical investigation of the second mechanism, which will require a much higher spatial resolution, will be undertaken soon.

Naturally all these processes require the existence of particle flows with reasonable amounts of kinetic energy. There are several recent publications [1--11] cataloguing  enough observational evidence  for such flows in the regions between the sun and the corona to warrant a serious investigation in this direction. It must be admitted  that we still have little understanding of the nature of the processes by which the relatively cool material (no hotter than about 20000K) moves upward from low altitudes (as low as a few thousand kilometers) to the outer atmosphere. For this paper, we shall simply exploit the observation that the flows exist, and work out their consequences. We believe that the flows might prove to be a crucial element in solving the riddle of 
coronal heating.

The model for the solar atmosphere that we propose and investigate is obtained  by injecting an essential new feature into several extant notions --- the plasma flows are allowed to play  their appropriate role in determining the evolution and the equilibrium properties of the structures under investigation. We reiterate that the distinguishing ingredient of our model is the assumption (observationally suggested) that relatively cold  particles spanning an entire range of velocity spectrum --- slow as well as fast,  continually flow from the sub--coronal to the coronal regions. It is the interaction of these cold primary  flows with the solar magnetic fields, and the strong coupling between the fluid 
and the magnetic aspects of the plasma that will define the characteristics 
of a typical coronal structure (including Coronal Holes). In this paper we
limit ourselves to the formation and primary heating aspects; we do not deal with instabilities, their nonlinear effects, flaring etc. These are the problems that we will confront at the next stage of the development of the model.

\bigskip

In Sec.2, we desribe in relative detail our basic  model for the upper solar
atmosphere, a time--dependent, two--fluid system of currents and flows. The flows are treated at par with other determining dynamical quantities, the currents and the solar magnetic fields. Section 3 is devoted to the derivation of the characteristics of typical coronal structures from the basic model. Following a general discussion, we numerically simulate the evolution of a cold plasma flow as it interacts with the solar magnetic field and gravity in Sec.3.1. We follow the fate of an initial cold supersonic flow as the particles get trapped by the magnetic field. By the time a sizeable density is built up we also find a considerable rise in temperature. In a very short time the velocity 
field developes a shocklike structure which dissipates with ordinary viscosity to convert the flow kinetic energy to heat. In Sec. 3.2 we take a different approach, and describe elements of the recently investigated magneto--fluid theory (see Mahajan and Yoshida, 1998, 2000) which allows the existence of equilibrium solutions missing in the flowless MHD. We find that a short--scale velocity component is predicted to be an essential aspect of a class of magnetofluid states in terms of which a typical coronal structure could be modelled. The magnetofluid states are the equilibrium states created by the strong interaction of the magnetic and the fluid character of a plasma, and are derived from the normal two--fluid equations when the velocity field is treated at par with the magnetic field. In a somewhat detailed discussion, we argue for the relevance of these states for the solar corona. These states could be seen as a set of quasi--equilibria evolving to an eventual hot coronal structure; the dissipation of the small scale velocity component provides the necessary source of heating. Since the numerical simulation of these states requires a much finer resolution than we have in our code, their time dependent simulation is deferred to a future work.

\section{THE BASIC MODEL --- GENERAL EQUATIONS FOR THE QUIESCENT SOLAR 
ATMOSPHERE}

In this section we will develop a general theoretical framework from which the typical solar coronal structure will be ``derived." In our model, the plasma flows from the Sun's surface provide the basic source of matter and energy for the myriad of coronal structures (including Coronal Holes). Although the magnetic field is, naturally, the primary culprit behind the structural diversity of the corona, the flows (and their interactions with the magnetic field) are expected to add substantially to that richness.

The primary objective of this paper is to investigate how these flows are trapped and heated in the closed magnetic field regions, and create one of the typical shining coronal elements. We shall, however, make a small digression to suggest a possible fate of the fast flows making their way through the regions where the magnetic field is weak, or has open field lines. The faster particles could readily escape the solar atmosphere in the open field-line regions. They could also do so  by punching temporary channels in the neighboring closed field--line  structures. The flows escaping through these existing or ``created" coronal  holes (the coronal holes (CH) are highly dynamical structures with open and ``nearly open"  magnetic field regions, see e.g. [17]) may eventually appear as the fast solar  wind.

In the closed field--line, the magnetic fields will trap the flows, and the trapping will lead to an accumulation of particles and energy creating the coronal elements with high temperature and density. We shall not consider the solar activity processes, since the activity regions (AR) and flares, though an additional source of particles and energy, cannot account for the continuous supply needed to maintain the corona. Moreover, in the theory we suggest, the flare is understood to be a secondary event and not the primary source for the creation of the hot corona.

To describe the entire atmosphere of the quiescent, non--flaring Sun we use the two--fluid equations where we keep the flow vorticity and viscosity effects (Hall MHD). The general equations will apply in both the open and the closed field regions. The difference between various sub--units of the atmosphere will come from the initial, and the boundary conditions.

\vspace{1cm}

Let ${\bf V}$ denotes the flow velocity field of the plasma in a region where the primary solar magnetic field is ${\bf B}_s$ . It is, of course, understood that the processes which generate the primary flows and the primary solar magnetic fields are independent (say at $t=0$ time). The total current ${\bf j}={\bf j}_f+{\bf j}_s$ (here ${\bf j}_f$ is the self--current that generates the magnetic field ${\bf B}_f$ and ${\bf j}_s$ is the source of the solar field ${\bf B}_s$) is related to the total (that can be observed) magnetic field ${\bf B}={\bf B}_s+{\bf B}_f$ by Amp\'ere's law:
$$
{\bf j}=\frac{c}{4\pi}\nabla\times{\bf B}. \eqno(1)
$$
Notice that in the framework we are developing (assumption of the existence of primary flows), the boundaries between the photosphere, the chromosphere and the corona become rather artificial; the different regions of each coronal structure are distinguished by just the parameters like the temperature and the density. In fact, these parameters should not show any discontinuities; they must change smoothly along the structure. At some distance from the Sun's surface, the plasma may become so hot and dense that it becomes visible (the bright, visible corona), and this altitude could be viewed as the base of the corona. But to study the creation and dynamics of bright coronal structures (loops, arches, arcades etc.) we must begin from the photosphere, and determine the plasma behavior in the closed field regions.

Assuming that the primary flows provide, on a continuous basis, the entire material for coronal structures, the solar flow with density $n$\ will obey the Continuity equation:
$$
\frac{\partial}{\partial t} n+\nabla\cdot(n{\bf V})=0.  \eqno(2)
$$
We must add a word of caution: in the closed field regions, the trapped particle density may become too high for the confining field, resulting in instabilities of all kinds. In this paper we shall not deal with instabilities and their consequences; it will constitute the next stage of development of the model.

Since the corona as well as the SW are known to be mostly hydrogen plasmas (with a small fraction of Helium, and neutrons, and an insignificant amount of highly ionized metallic atoms) with nearly equal electron and proton densities: $n_e\simeq n_i=n$ , we expect the quasineutrality condition $\nabla\cdot{\bf j}=0$ to hold.

In what follows, we shall assume that the electron and the proton flow
velocities are different (two--fluid approximation was used e.g. in Sturrock and Hartly, (1966). Neglecting electron inertia, these are ${\bf V}_i={\bf V}$, and ${\bf V}_e=({\bf V}-{\bf j}/en)$, respectively. We assign equal temperatures to the electron and the protons for processes associated with the quiescent Sun. For the creation processes of a typical coronal structure, this assumption is quite good. For the fast SW, however, we know from recent observations (Banaszkiewicz {\it et~al.} 1997 and references therein),  that the species temperatures are found to be different: $T_i\sim 2\cdot 10^5\,$K and $T_e\sim 1\cdot 10^5\,$K. Since  the fast SW is not the principal interest of this paper, we shall persist with the equal temperatures assumption; the kinetic pressure $p$ is given by:
$$
p=p_i+p_e\simeq 2\,nT,\qquad T=T_i\simeq T_e. \eqno(3)
$$
With this expression for $p$, and by neglecting electron inertia, the two--fluid equations  are obtained by combining the proton and the electron equations of motion:
$$\frac{\partial}{\partial t} V_k+({\bf V}\cdot\nabla)V_k=$$
$$
=\frac{1}{en}({\bf j}\times{\bf b})_k-\frac{2}{nm_i}\nabla_k(nT)
+\nabla_k\left(\frac{M_\odot
G}{r}\right)-\frac{1}{nm_i}\nabla_l\Pi_{i,kl},  \eqno(4)
$$
and
$$
\frac{\partial}{\partial t}{\bf b}-\nabla\times\left[\left({\bf
V}-\frac{\bf j}{en}\right)\times{\bf b}\right]
=\frac{2}{m_i}\nabla\left(\frac{1}{n}\right)
\times\nabla(nT), \eqno(5)
$$
where \ ${\bf b}={e\bf B}/{m_ic}$, \ $m_i$ is the proton mass, $G$ is the gravitational constant, $M_\odot$ is the solar mass, $r$ is the radial distance, and $\Pi_{i,lk}$ is the ion viscosity tensor. For flows with large spatial variation, the viscous term will end up playing an important part. To obtain an equation for the evolution of the flow temperature $T$, we begin with the energy balance equations for a magnetized, neutral, isothermal electron--proton plasma:
$$
\frac{\partial}{\partial t}\varepsilon_\alpha+\nabla_k(\varepsilon_\alpha
V_{\alpha,k}+{\cal P}_{\alpha,kl} V_{\alpha,l})+\nabla {\bf q}_\alpha=
n_\alpha{\bf f}_\alpha\cdot{\bf V}_\alpha,  \eqno(6)
$$
where $\alpha$ is the species index. The fluid energy $\varepsilon_\alpha$
(thermal energy $+$ kinetic energy) and the total pressure tensor 
${\cal P}_{\alpha,kl}$ are given by
$$
\varepsilon_\alpha=n_\alpha\left(\frac{3}{2} T_\alpha+\frac{m_\alpha
V_\alpha^2}{2}\right),\qquad{\cal P}_{\alpha,kl}=n_\alpha
T_\alpha\delta_{kl}+\Pi_{\alpha,kl},  \eqno(7)
$$
and
$$
{\bf f}_\alpha=e_\alpha{\bf E}+\frac{e_\alpha}{c}{\bf V}_\alpha\times
{\bf B}+m_\alpha\nabla\frac{GM_\odot}{r},  \eqno(8)
$$
is the volume force experienced by the fluids (${\bf E}$ is the electric field). In Eq.~(6), ${\bf q}_\alpha $ is the heat flux density for the species $\alpha $. After standard manipulations we arrive at the temperature evolution equation
$$
\frac{3}{2} n\frac{d}{dt}(2T)+\nabla({\bf q}_i+{\bf q}_e)= -2nT\nabla
\cdot {\bf V}+ m_in\nu_i\left[\frac{1}{2}\left(\frac{\partial 
V_k}{\partial
x_l}+\frac{\partial V_l}{\partial x_k}\right)^2-\frac{2}{3} 
(\nabla\cdot
{\bf V})^2\right]+
$$
$$
+\frac{5}{2} n\left(\frac{\bf j}{en}\cdot\nabla T\right)-\frac{\bf 
j}{en}\nabla(nT)+E_H+E_R\ \eqno(9)
$$
where $E_R$ is the total radiative loss, $E_H$ is the local mechanical
heating function, and $\nu_i$ is the ion kinematic viscosity. Note that we have retained viscous dissipation in this system. If primary flows are ignored in the theory, various anomalous heating mechanisms need to be invoked, and a corresponding term $E_H$ has to be added. The full viscosity tensor relevant to a magnetized plasma is rather cumbersome, and we do not display it here. However just to have a feel for the importance of spatial variation in viscous dissipation, we display its relatively simple symmetric form. It is to be clearly understood that this version is meant only for theoretical elucidation and not for detailed simulation. We notice that even for incompressible and 
currentless flows, heat can be generated from the viscous dissipation of the flow vorticity. For such a simple system, the rate of kinetic energy 
dissipation turns out to be
$$
\left[\frac{d}{dt}\left(\frac{m_i{\bf V}^2}{2}\right)\right]_{\rm visc}
=-m_in\nu_i\left(\frac{1}{2}(\nabla\times{\bf V})^2+\frac{2}{3}(\nabla
\cdot {\bf V})^2\right).  \eqno(10)
$$
revealing that for an incompressible plasma, the greater the vorticity 
of the flow, the greater the rate of dissipation.

Let us now introduce the following dimensionless variables:
$$
{\bf r}\to{\bf r}\ R_\odot; \qquad t\to t\ \frac{R_\odot}{V_A}; \qquad
{\bf b}\to{\bf b}\ b_\odot; \qquad T\to T\ T_\odot; \qquad n\to n\
n_\odot;
$$
$$
{\bf V}\to {\bf V}\ V_A; \qquad {\bf j}\to {\bf j}\ V_Aen_\odot;
\qquad{\bf q}_\alpha
\to{\bf q}_\alpha n_\odot T_\odot V_A; \qquad \nu_i\to\nu_i\
R_\odot V_A \eqno(11a),
$$
and parameters:
$$
b_\odot=\frac{eB(R_\odot)}{m_ic};
\qquad\lambda_{i\odot}=\frac{c}{\omega_{i\odot}};\qquad
c_s^2=\frac{2T_\odot}{m_i};
\qquad\omega_{i\odot}^2=\frac{4\pi e^2n_\odot}{m_i}; \qquad
V_A=b_\odot\lambda_{i\odot};
$$
$$
r_A=\frac{GM_\odot}{V_A^2R_\odot}=2\beta\ r_c;\qquad r_c=\frac{GM_\odot}
{2c_s^2R_\odot};\qquad \alpha=\frac{\lambda_{i\odot}}{R_\odot};\qquad
\beta=\frac{c_s^2}{V_A^2}, \eqno(11b)
$$
where $R_\odot$ is the solar radius. Note that in general $\nu_i$ is a
function of density and temperature: $\nu_i=(V_{i,th}T^2/12\pi ne^4)$.

In terms of these variables, our equations read:
\setcounter{equation}{11}
$$
\frac{\partial}{\partial t}{\bf V}+({\bf V}\cdot\nabla){\bf V}=
$$
\bq
=\frac{1}{n}\nabla\times{\bf b}\times {\bf b}
-\beta\frac{1}{n}\nabla(nT)+\nabla\left(\frac{r_A}{r}\right)+\nu_i\left(
\nabla^2{\bf V}+\frac{1}{3}\nabla(\nabla\cdot{\bf V})\right), 
\eq
% \eqno(12)
\bq
\frac{\partial}{\partial t}{\bf b}-\nabla\times\left({\bf V}-\frac{\alpha}{n}
\nabla\times {\bf b}\right)\times{\bf b}=\alpha\beta\
\nabla\left(\frac{1}{n}\right)\times \nabla (nT),
\eq
% \eqno(13)
\bq
\nabla\cdot{\bf b}=0,
\eq
% \eqno(14)
\bq
\frac{\partial}{\partial t} n+\nabla\cdot n{\bf V}=0,
\eq
% \eqno(15)
$$
\frac{3}{2} n\frac{d}{dt}(2T)+\nabla ({\bf q}_i+{\bf q}_e)= 
-2nT\nabla
\cdot{\bf V}+2\beta^{-1}\nu_in\left[\frac{1}{2}\left(\frac{\partial
V_k}{\partial x_l}
+\frac{\partial V_l}{\partial x_k}\right)^2-\frac{2}{3}(\nabla\cdot
{\bf V})^2\right]+
$$
$$
+\frac{5}{2}\alpha(\nabla\times {\bf b})\cdot\nabla T-\frac{\alpha}{n
}(\nabla \times {\bf b})\nabla(nT)+E_H+E_R.\eqno(16)
$$
This set of equations will now be studied for different types of magnetic
field regions, in particular the regions with closed field lines.

\bigskip

Before we embark on a detailed theory of the formation and heating of the corona, we would like to give a short list of heating mechanisms which have been invoked to deal with this rather fundamental and still unresolved problem of Solar physics : Alfv\'en waves [20--27], Magnetic reconnection in Current sheets [28--36], and MHD Turbulence [37--39]. For all these schemes, the predicted temperature profiles in the coronal structures come out to be highly sensitive to the form of the heating mechanism [40,41]. Parker (1988) suggested that the solar corona could be heated by dissipation of many tangential discontinuities arising spontaneously in the coronal magnetic field that is stirred by random photospheric footpoint motions. This theory stimulated numerous searches for observational signatures of nanoflares. Unfortunately, all of these attempts fall short of providing a continuous (both in space and time) energy supply that is required to first create in a few minutes, and then support for longer periods the observed bright coronal structures (see e.g. [1,2]).

Our attempt to solve this problem makes a clean break with the conventional approach. We do not look for the energy source within the corona but place it squarely in the primary flows emerging from the Sun (see the results of [1--3]). We propose (and will test) the hypothesis that the energy and particles associated with the primary flows, in interaction with the magnetic field, do not only create the variety of configurations which constitute the corona, but also provide the primary heating. The flows can give energy and particle supply to these regions on a continuous basis --- we will show that the primary heating takes place simultaneously with the accumulation of the corona and a major aspect of the flow--magnetic field interaction, for our system, is to provide a pathway for this to happen.

A mathematical modeling of the coronal structure (for its creation and primary heating) will require the solution of Eqs.~(12)--(16) with appropriate initial and boundary conditions. We will use a mixture of analytical and numerical methods to extract, what we believe, is a reasonable picture of the salient aspects of a typical coronal structure.

\vspace{1cm}

%sec.III

\section{CONSTRUCTION OF A TYPICAL CORONAL STRUCTURE}

Though the solar atmosphere is highly structured, it seems that most of the constituent elements have something common in their creation and heating. In order to construct a unified theory for the entire corona, one would have to confront large variations in plasma density and temperature. It seems, however, that beyond the coronal base, the equilibrium temperature tends to be nearly constant on each one of these structures; the temperature of a specific structure increases insignificantly (about 20 p.c.) from its value at the base to its maximum reached at the top of the structure. This change is much less than the temperature  change (about 2 orders of magnitude) that occurs between 
the solar surface and the coronal base. This observation is an outcome of the investigation of several authors (see, for example, [1,2,41--45]). Their results show that the bright elements of the corona are composed of quasi--isothermic and ultra--thin arcs (loops) of different temperature and density, situated (located) close to one other. This state is, perhaps, brought about by the isolating influence of magnetic fields which prevent the particle and energy transfer between neighboring structures. 

It is safe to assume, then, that in the quasi--equilibrium state, each coronal structure has a nearly constant temperature, but different structures have different characteristic temperatures, i.e., the bright corona seen as a single entity will have considerable temperature variation. Observations tell us that the coronal temperatures are much higher than those of the primary flows (which we are proposing as the mother of the corona). For the consistency of the model, therefore, it is essential that the primary ``heating" must take place during the process of  accumulation of a given coronal entity.

This apparent problem, in fact, can be converted to a theoretical advantage. We distinguish two important eras in the life of a coronal structure; a hectic period when it acquires particles and energy (accumulation and heating), and the relatively calmer period when it "shines" as a bright, high temperature object.

In the first era, the most important issue is that of heating while particle accumulation (trapping) takes place in a curved magnetic field. This is, in fact, the essential new ingredient of the current approach. We plan to show:

1) that the kinetic energy contained in the primary flows can be dissipated by viscosity to heat the plasma, and 
2) that this dissipation can be large enough to produce the observed
temperatures.

Naturally, a time dependent treatment will be needed to describe this era.

Any additional heating mechanisms, operative after the emergence of the coronal structure, will not be discussed in this paper. For an essential energy inventory of the quasi-equilibrium coronal structure, we also ignore the contributions of flares and other ``activities" on the solar surface because they do not provide a continuous and sufficient energy supply [2].

The second era is that of the quasi-equilibrium of a coronal structure of
given density and temperature - neither of which has to be strictly constant. The primary heating has already been performed, and in the equilibrium state, we can neglect viscosity, resistivity and other collisional effects in addition to neglecting the time dependence. The calculations in this regime will be limited to the determination of the magnetic field and the velocity--field structures that the collisionless magnetofluids can generate and we will also examine if these structures can confine plasma pressure.

%subsec 3.1

\subsection{Creation and heating of coronal structure}

In this subsection we will concentrate on numerical methods to test our basic conjecture that the primary solar flows are responsible for the creation and heating of  a typical bright coronal structure. The numerical results (obtained by modeling Eqs.~(12)--(16) with viscosity tensor relevant to magnetized plasma) are extremely preliminary, but they clearly indicate that the proposed mechanism has considerable promise.

Let us first make order of magnitude estimates on the requirements that must be met for this scheme to be meaningful. It is well known that (see e.g. [46]) 
the rate of energy losses $F$ from the solar corona by radiation, thermal conduction, and advection is approximately $5\cdot 10^5\, \mbox{erg/cm}^2\,$s. For the brightest loops the rate loss could even reach $5\cdot10^6\,\mbox{erg/cm}^2\,$s. If the conversion of the kinetic energy in the primary flows were to compensate for these losses, we would require a radial energy flux 
\setcounter{equation}{16}
$$
\frac{1}{2}m_in_0V_0^2\ V_0\ge F,\eqno(17)
$$
where $V_0$ is the initial flow speed. For $V_0\sim 300\,$km/s this implies an initial density in the range: $(3\cdot 10^7 - 4\cdot 10^8\,)\mbox{cm}^{-3}$.

For slower ($\sim 100\,$km/s) velocity primary flows the starting density has to be higher ($\ge 10^9\,\mbox{cm}^{-3}$). These values seem reasonable according to the latest observational data [1-3].

The normal viscous dissipation of the flow takes place on a time (using Eq.~(10)):
$$
t_{{\rm visc}}\sim\frac{L^2}{\nu_i},\eqno(18)
$$
where $L$ is the length of the coronal structure. For a primary flow with $T_0=3\,\mbox{eV}= 3.5\cdot 10^4\,$K and  $n_0=4\cdot 10^8\,\mbox{cm}^{-3}$ creating a quiet coronal structure of size $L=(2\cdot 10^9 - 7\cdot 10^{10})\,\mbox{cm}$, the dissipation time can be estimated to be of the order of $(2\cdot 10^8 - 10^{10})\,$s. The shorter the structure and hotter the flow, the faster is the rate of dissipation. This estimated time is much longer than what is actually found by the latest observations by TRACE [1].  Mechanisms much
faster than the one embodied in (18), therefore, will be needed for the model to work. In the absence of ``anamalous viscosity", the only way to enhance the dissipation rates (to the observed values) is to create spatial gradients of the velocity field that are on a scale much much shorter than that of the structure length (defined by the smooth part of the magnetic field). Thus, the viability of the model depends wholly on the existence of mechanisms  that induce short--scale velocity fields. Numerical simulations show that the short--scale velocity fields are, indeed, self--consistently generated in the two--fluid system.

For numerical work (to illustrate the bright coronal structure formation), we model the initial solar magnetic field as a 2D arcade with circular field lines in the $x$--$z$~plane (see Fig.~1 for the contours of the vector potential, or the flux function). The field attains its maximum value  $B_{\rm max}(x_o, z=0)$ at $x_0$ at the center of the arcade, and is a decreasing function of the height $z$ (radial direction). The set of  model equations (12-16) was solved in 2D flat geometry (x,z) using the 2D version of Lax--Wendroff numerical scheme (Richtmyer and Morton 1967) alongwith applying the Flux--Corrected--Transport procedure [48]. Equation (13) was replaced with its equivalent for the y--component of the vector potential which automatically ensures the divergence-free property of the magnetic field. The equation of heat conduction was treated separately by Alternate Direction Implicit method with iterations [49].
Transport coefficients for heat conduction and viscosity were taken from Braginski, 1965. A numerical mesh of \ $200\times 150$\  points was used for computation.

To illustrate the formation and heating of a general coronal structure, we have modeled several cases with different initial and boundary conditions for cold primary flows. The dynamical picture is strongly dependent on the relation of the initial flow pressure and  the magnetic field strength. Two limiting cases are interesting: 1) the initial magnetic field is weak, and the flow significantly deforms (and in specific cases, drags) the magnetic field lines, 2) the initial magnetic field is strong, and the flow leaves the field lines  practically unchanged.

For sub--Alfv\'enic flows, we present in Figs. 2-5 \ the salient features of our preliminary results. We have plotted (as functions of $x$ and $z$) four relevant physical quantities: the flux function \ $A$, the density \ $n$, the temperature \ $T$, and the magnitude of the velocity field \ $|{\bf V}|$ (for specific cases, when needed, we  give  the radial component of velocity field \ $V_z$\ also).

The plots correspond to two (in some cases to three) different time frames.
The results are described under three separate  headings, covering respectively, the fully uniform, the spatially non--uniform, and the time--dependent as well as spatially non--uniform initial flows.

\bigskip

A. {\bf Initially uniform primary flow and an Arcade-like magnetic field structure} --- Fig.~2.

This case is highly idealized but illustrates the main aspects of the
creation of the  hot coronal structures, and of the basic heating process.

When discussing the temporally uniform initial flows, we choose the 
parameters to satisfy the observational constraint that, over a period of 
some tens of minutes, the location of the heating must have a relatively smooth
evolution  [1]. The final shape and location of the coronal structure (of the associated ${\bf B}({\bf r},t)$, for example) will be naturally defined by its 
material source, by the  heating dynamics, and   by the  initial field ${\bf B}_0({\bf r},t)$.

For these studies, the initial flow velocity field is taken to be uniform at the surface and has only  a radial component, $V_z=300\,km/s$. Other parameters are: Maximum value of the magnetic field $B_{\rm max}(x_o, z=0)=7G$, initial density of the flow  $4\cdot 10^8\,cm^{-3}$ and the initial temperature $3\,eV$. Simulations yield the following results:

1) The flow particles begin to accumulate at the footpoints near the solar
surface (Fig.~2, see density at $t=750\,$s). The accumulation goes on with time, and gradually the entire volume under the arcade (starting from the central short loops) is filled with particles (Fig.~2, density at $t=1400\,$s). First the shorter loops are filled, and then the larger ones.

2) The heating of the particles goes hand in hand with the accumulation
(Fig.~2, plots for density and temperature).

3) The regions of stronger magnetic fields are denser in population 
(Fig.~2, plots for $A$ and $n$). In earlier stages of the formation of a 
coronal structure, the regions near the base (where the field is stronger) 
are denser and hotter than the distant regions (Fig.~2, $t=750\,$s \ 
plots for $n$ and $T$\ ); for shorter loops, the density increases (as a function of height $z$) from the bottom of the structure, and then falls --- first rapidly and later insignificantly; the maximum density is much greater than the initial density of the flows.

4) The dissipation of the flow kinetic energy is faster in the first stage
of formation (Fig.~2, $t=750\,$s plot for $|V|$ ). The plot $|V|$ versus
$z$ shows steep (shock--like) gradients  near the base. Thus the bright
base is created in the very first stage in the stronger magnetic field
regions (shorter loops). For given parameters, the initial flow is strongly supersonic. Thus the shocks are generated with efficient transfer of kinetic energy into heat. As the mean free path of ions in the plasma is of the order of $(10^6 - 10^7\,)cm$ (in the direction parallel to the magnetic field) and the dimension of the structure is much greater -- of the order of $10^{10}\,cm$  -- efficient conditions for the kinetic energy dissipation exist. The plots for the velocity, temperature and density reveal that with increasing z, and in the regions  away from the arcade center, we first find an undisturbed flow with low temperature, then see  a transient area with high density and temperature, and finally  a shock consistent with Hugoniot conditions. The short scale represented by the width of the shock-layer (determined by viscosity) is the main enhancer of  viscous dissipation.

5) For later times, the brightening process spreads over wide regions (Fig.~2, $t=1400\,$s plot for temperature).

6) In the very first stage, the shorter loops are a bit overheated, but they cool down somewhat at later times when the longer loops begin to get heated (Fig.~2, plots for temperature).

7) The base ($T\ge 100\,$eV) of the bright region is at about $1.4\cdot 10^{9}\,cm\sim 0.02R_\odot$ (Fig.~2, $t=1400\,s$ plots for $n$ and $T$) from the solar surface. This number is in a very good agreement with the latest TRACE results [1]. Outwards from the base, the accumulated layer has somewhat lower, but more or less uniform, insignificantly decreasing density. In the accumulated layer the kinetic energy of the flow is essentially uniform (again, decreases insignificantly); the dissipation has practically stopped (Fig.~2, $t=1400\,s=23\,min$\ , plot for $|V|$ versus $z$). The temperature is practically uniform in the longer loops and increases insignificantly in shorter loops (for some special conditions these conclusions may be somewhat modified in specific regions of the arcade; see point 8)\,). Outwards from the hottest region of the arcade, the temperature decreases gradually and at some radial distance the outer boundary of the bright part is reached (Fig.~2, $t=1400\,s$ plot for temperature). Thus, in a very short time a dense and bright ``coronal structure" is created --- this object survives for a time much longer  than was needed for its creation. The simulations show that the heating process may continue during this so--called equilibrium stage, but at a rate much slower than the earlier primary heating. This heating seems just additional and supporting to the heat content of the nascent hot structure. At this time, however, the velocity field is already much smaller in magnitude as compared to the initial values; the flows in the hot coronal structure are already  subsonic. This is a possible explanation why supersonic flows may not be seen in the hot observable coronal structures.

8) When relatively  dense primary flows interact with weak arcade-like magnetic fields ($B_{\rm max}(x_0,z_0=0)\leq 10\,G$ for our initial flow with given above parameters), the field lines begin to deform (soon after the creation of the solar base) in the central region of the arcade but far from the base (see $t=1400\,s$ plots for density and temperature in Fig.~2). The particle accumulation is still strong, and the dissipation, though quite fast, stops rather rapidly. Consequently, the temperature first reaches a maximum (up to the deformed field--line region this maximum is reached at the summit for each short loop) and later falls rapidly.  Gradually one can see signs for the creation of a local gravitational potential well behind the shortest loops (see $t=1400\,$s plot for $A$ in Fig.~2). This well supports a relatively dense and cold plasma in the central area of the arcade ($t=1400\,s$ for $n$ and $T$ of Fig.~2). The density of this  structure is considerably greater than that of the surrounding areas, and the temperature is considerably less than that of the rest of the accumulated regions at the same height of the arcade.

Our preliminary simulations show that for the same parameters of the primary outflow, such cold and dense plasma objects (confined in the so--called potential well) will not form in the regions where the initial  magnetic fields are stronger ($B_{\rm max}(x_0,z_0=0)\geq 20\,G$).

\bigskip
B. {\bf Spatially non--uniform primary flow interacting with an arcade--like
magnetic field structure} --- Figs. 3, 4.

The latest observations support the idea that the coronal material is injected discontinously (in pulses or bunches, for example) from lower altitudes into the regions of interest  (e.g. spicules, jet--like structures [6,7,12,13,1,2]). A realistic simulation, then, requires a study of the interaction of spatially 
non--uniform initial flows with arcade--like magnetic field structures. These ``close to the actual'' cases represent more vividly the dynamics of the hot coronal formation.

1) When the spatially symmetric initial flow (plot for $V_z$ at $t=0\ $ in Fig.~3a) interacts with the arcade (plots in Fig.~3), and the initial magnetic field is rather strong ($B_{\rm max}(x_0,z_0=0)=20\,G$), the primary heating is
completed in a very short time ($\sim (2-3) min $) on  distances ($\sim 10000\,km$) shorter than the uniform--flow case when the initial magnetic field was weaker. This is also consistent with observations.  The heating is very symmetric and the resulting hot structure is uniformly heated to $1.6\cdot 10^6 K$.

2) Observations reveal the existence of cool material and downflows, right within the hot coronal structures; they also show an imbalance in the primary heating on the two sides of the loops (see [12, 1]). To reproduce these characteristics, we have modeled the coronal structure formation process using an asymmetric, spatially non--uniform initial flow interacting with a strong magnetic field (see Fig.~4).  

For both of the  discussed cases, the downflows can be clearly seen for the 
velocity field component $V_z$. In Fig.~4, the downflow is created simply by changing the initial character of the flow (initially we had only the right pulse from the velocity field distribution given in  Fig.~3a), while in Fig.~3a (plot at $t=297\,s$), the downflows are the result of more complicated events (see explanation below, in the next paragraph). The final parameters of the downflows are strongly dependent on the initial and boundary conditions. In the pictures, the imbalance in the primary heating process is also revealed.

When two identical pulses (Fig.~3a, plot at $t=0\,s$) enter in succession into our standard, arcade--like initial magnetic field, we simulate the equivalent of 
two colliding flows on the top of a structure. Shocks, though not very strong, 
are generated in a very short time ($t=30\,s$). Such shocks, on both sides of the arcade--center, have hot fronts and cold tails. Soon ($t=42\,s$) these shocks become "visible", a hot and dense area is created on  top of the structure where these shocks (at this moment they have become stronger) collide. After the collision (and "reflection"), the entire area within the arcade becomes gradually hot. At some moment, a practically uniformly heated structure is created, and the primary heating  stops. This process is accompanied by
downflows much slower than the primary flows; much of the primary flow
kinetic energy has been converted to heat via shock generation (the shock and downflow velocities differ significantly). It is clear that in the case of spatially assymetric initial flows, the downflows on different sides of the arcade--center will have different characteristics. Due to the high pressure prevalent in the nascent hot structure (loop), there is no more inflow of the plasma and the flow deposits its energy at the base; the base becomes overheated. Later this energy can be again transferred upwards via thermal conduction (this mechanism can work in all the discussed cases), but at that moment the flow could be also changed (see initially time--dependent flow cases below).

Plots for the temperature and velocity field in Figs.~3b,\,4\  also indicate that some cold particles still remain in the body of the newly created hot structure. These particles are perhaps from the slower aggregates (our initial flow was not uniform) which did not have sufficient energy to be converted to heat.

\bigskip

C. {\bf  Time dependent non--uniform initial flows interacting with
arcade--like magnetic field structures.} -- Fig.~5

To simulate  reality further we introduce time dependence in the initial primary flow velocity field. We discuss two distinct cases:

1) Initially, the velocity field has a pulse--like distribution with a time--period  nearly half of the ``formation time'' of the quasi-equilibrium structure corresponding to the case with time--independent initial conditions. The results displayed in Fig.~5 show that the emerging coronal structure has a rather uniform distribution of temperature along the magnetic field, and the latter is practically undeformed during formation and heating. We see that when the basic heating  ceases, the hot structure survives for the time of computation which happens to be shorter than the time necessary for losses that destroy the structure.

2) The velocity field has a fast amplitude modulation near its maximum value (
for these simulations the maximum radial velocity was taken to be $300\,km/s$). We find that the dynamics of the hot coronal structure creation is quite similar to the initially time--independent, spatially symmetric case. Because of this, we don't give here the corresponding plots. We only note that for this case, the structure tends to become even hotter (by a factor \,$1.2$\, for the same parameters) and when quasi--equilibrium is established (time for this to happen is longer than for the time--independent initial flows) the base of the structure is  hotter than the top although at an earlier time the top was hotter, i.e, there is a temperature oscillation with a time--period longer compared to the creation time of the hot structure.

\bigskip

The main message of numerical simulation is that the dynamical interaction of an initial flow with the ambient solar magnetic field leads to a re--organization of the plasma such that the regions in the close vicinity of the solar surface are characterized by strongly varying (in space and time) density and temperature, and even faster varying velocity field, while the regions farther out from the bright base are nearly uniform in these physical parameters. This phenomenon pertains generally, and not for just a set of specific structures. The creation and primary heating of the coronal structures are simultaneous, accompanied by strong shocks. These are fast processes (few tens of minutes) taking place at very short radial distances from the Sun ($\sim 10000\,km$) in the strong magnetic field regions with significant curvature. The final 
characteristics of the created coronal structures are defined by the boundary conditions for the coupled primary flow--solar magnetic field system. The stronger the magnetic field, the faster is the process of creation of the hot coronal structure with its base nearer the solar surface. To investigate the near surface region one must use general time--dependent 3D equations. Quasi--stationary (equilibrium) equations, on the other hand, will suffice to describe the hot and bright layers --- the already existing visible coronal structures.

\bigskip

\subsection{Construction of quasi--equilibrium coronal structure}

The familiar MHD theory (single--fluid) is a reduced case of the more general two--fluid theory discussed in this paper. Constrained minimizaion of the magnetic energy  in MHD leads to force--free static equilibrium configurations [50,51]. The range of two--fluid relaxed states, however, is considerably  larger because the velocity  field, now, begins to play an independent fundamental role. The presence of the velocity field not only leads to new pressure confining states [52,53], but also to the possibility of heating the equilibrium structures by the dissipation of kinetic energy. The latter feature is highly desirable if these equilibria were to be somehow related to the bright coronal structures.

We begin investigating the two--fluid states by first studying the simplest, almost analytically tractable, equilibria. This happens when the pressure term in the equation of motion (12) becomes a full gradient, i.e, whenever an equation of state relating the pressure and density can be invoked. For our present purpose, we limit ourselves to the constant temperature states allowing 
\ $n^{-1}\nabla p\to 2\,T\nabla\ln \ n$ \ . \

Normalizing $n$ to some constant coronal base density $n_0$ (reminding the reader that $n_0$ is different for different structures!), and using our other standard normalizations ($\lambda_{i0}=c/\omega_{i0}$ is defined with $n_0$), our system of equations reduces to:
$$
\frac{1}{n}\nabla\times {\bf b\times b}+\nabla\left(\frac{r_{A0}}{r}
-\beta_0\ln\ n-\frac{V^2}{2}\right)+ {\bf V\times (\nabla\times
V})=0, \eqno(19)
$$
$$
\nabla\times\left({\bf V}-\frac{\alpha_0}{n}\nabla \times {\bf b}
\right)\times {\bf b}=0, \eqno(20)
$$
$$
\nabla \cdot (n{\bf V})=0,  \eqno(21)
$$
where $r_{A0}, \ \alpha_0, \ \beta_0$ are defined with $n_0, \ T_0, \ B_0$.
This is a complete system of seven equations in seven variables.

Following Mahajan and Yoshida (1998) and [54], we seek equilibrium solutions of the simplest kind.  Straightforward algebra leads us to the following system of linear equations:
$$
{\bf b}+\alpha_0 \nabla\times {\bf V}=d\ n\ {\bf V} \eqno(22)
$$
and
$$
{\bf b}=a\ n\ \left[{\bf V}-\frac{\alpha_0}{n}\,\nabla\times {\bf
b}\right], \eqno(23)
$$
where $a$ and $d$ are dimensionless constants related to the two invariants: the magnetic helicity \ $\int ({\bf A}\cdot {\bf B})\ d^3x $ \  and the generalized helicity \ $\int ({\bf A}+{\bf V})\cdot ({\bf B}+\nabla\times {\bf V})d^3x$  \ (or $\int ({\bf V}\cdot {\bf B}+{\bf A}\cdot \nabla \times {\bf V} +{\bf V}\cdot \nabla \times {\bf V})\ d^3x$ \ )  of the system. We will discuss $a$ and $d$ later. The equilibrium solutions (22), (23) encapsulate the simple physics: 1) the electrons follow the field lines, 2) while the ions, due to their inertia, follow the field lines modified by the fluid vorticity. These equations, when substituted in (19), (20), lead to 
$$
\nabla\left(\frac{r_{A0}}{r}-\beta_0\ln\
n-\frac{V^2}{2}\right)=0,\ \eqno(24)
$$
giving the Bernoulli condition which will determine the density of the
structure in terms of the flow kinetic energy, and solar gravity.
Equations~(22) and (23) are readily manipulated to yield
$$
\frac{\alpha_0^2}{n}\nabla\times\nabla\times{\bf V}+\alpha_0\
\nabla \times \left(\frac{1}{a}-d\ n\right){\bf V}+\left(1-\frac{d}{a}
\right){\bf V}=0. \eqno(25)
$$
which must be solved with (24) for $n$ and ${\bf V}$; the magnetic field can, then, be determined from (22).

Equation~(24) is solved to obtain
$$
n=\exp\left(-\left[ 
2g_0-\frac{V^2_0}{2\beta_0}-2g+\frac{V^2}{2\beta_0}
\right]\right), \eqno(26)
$$
where $g(r)=r_{c0}/r$. This relation is rather interesting; it tells us that the variation in density can be quite large for a low $\beta_0$ plasma (coronal plasmas tend to be low $\beta_0 $; the latter is in the range $0.004 - 0.05$) if the gravity and the flow kinetic energy vary on length scales comparable to the extent of the coronal structure. In this system of equations, as we mentioned above, the temperature (which defines $\beta_0$) has to be fixed by initial and boundary conditions at the base of the structure. Substituting (26) into (25) will yield a single equation for velocity which is quite nontrivially nonlinear. Numerical solutions of the equations are tedious but straightforward.

For analytical progress, essential to revealing the nature of the self--consistent fields and flows, we will now make the additional simplifying assumption of constant density. This is a rather drastic step (in numerical work, we take the density to be a proper dynamical variable) but it can help us a great deal in unraveling the underlying physics. There are two entirely different situations where this assumption may be justified:

1) the primary heating of corona has already been performed, i.e., a
substantial part of flow initial kinetic energy has been converted to 
heat. The rest of the kinetic energy, i.e., the kinetic energy of the 
equilibrium coronal structure is not expected to change much within the 
span of a given structure. Note that the ratio of velocity components will 
have a large spatial variation, but the variation in $V^2$ is expected to 
be small. It is also easy to estimate that within a typical structure, 
gravity varies quite insignificantly. There will be exceptional cases 
like the neighborhood of the Coronal holes and the streamer belts, where 
significant heating could still be going on, and the temperature and density variations could not be ignored. Such regions are extremely hard to model;

2) if the rates of kinetic energy dissipation are not very large, we can imagine the plasma to be going through a series of quasi--equilibria before it settles into a particular coronal structure. At each stage we need the velocity fields in order to know if an appropriate amount of heating can take place. The density variation, though a factor, is not crucial in an approximate estimation of the desired quantities.

The constant density assumption $n=1$ will be used only in Eq.~(25) to solve
for the velocity field (or the ${\bf b}$ field which will now obey the 
same equation). These solutions, when substituted in Eq.~(26), would determine the density profile (slowly varying) of a given structure.

In the rest of this sub--section we will present several classes of the
solutions of the following linear equation:
$$
\alpha_0^2\ \nabla\times \nabla \times {\bf Q}+ \alpha_0\ \left(\frac{1}{a}
-d\right)\nabla \times {\bf Q}+ \left(1-\frac{d}{a}\right){\bf Q}=0,  \eqno(27)
$$
where ${\bf Q}$ is either ${\bf V}$ or ${\bf b}$. To make contact with existing literature, we would use ${\bf b}$ as our basic field to be determined by Eq.~(25); the velocity field ${\bf V}$ will follow from Eqs.~(22) and (23), which for $n=1$, become 
$$
{\bf b}+\alpha_0 {\nabla}\times {\bf V}=d{\bf V} \eqno(22')
$$
and
$$
{\bf b}=a\left[{\bf V}-\ \alpha_0 \,\nabla\times {\bf b}\right].
\eqno(23')
$$

It is worth remarking that in order to derive the preceding set of equations, all we need is the constant density assumption; the temperature can have gradients and, these are determined from the Bernoulli condition (20) with $\beta_0(T)$ replacing $\beta_0\ln\ n$.

\subsection{Analysis of the $Curl\ Curl$ Equation, Typical Coronal
Equilibria}

The Double $Curl$ equation (27) was derived only recently [53] (Mahajan and Yoshida 1998); its potential, is still, largely unexplored(see [53], [54]). 
The extra double $curl$ (the very first) term distinguishes it from the standard force-free equation [55,50,56] (Woltjer 1958; Taylor 1974, 1986; Priest 1994 and references therein) used in the solar context. Since $a$ and $d$ are constants, Eq.~(25), without the double $curl$ term, reproduces what has been called the ``relaxed state" [50,56]. We will see that this term contains quantitative as well as qualitative new physics.

In an ideal magnetofluid, the parameters $a$ and $d$ are fixed by the initial conditions; these are the measures of the constants of motion, the magnetic helicity, and the fluid plus cross helicity or some linear combination thereof [53,52,57,38]. In our calculations, $a$ and $d$ will be considered as given quantities. The existence of two, rather than one (as in the standard relaxed equilibria) parameter in this theory is an indication that we may have, already, found an extra clue to answer the extremely important question: why do the coronal structures have a variety of length scales, and what are the determinants of these scales?

We also have the parameter $\alpha_0$, the ratio of the ion skin depth to
the solar radius. For typical densities of interest $(\sim (10^7-10^9\,) \mbox{cm}^{-3})$, its value ranges from ($\sim 10^{-7}-10^{-8}  $); a very small number, indeed. Let us also remind ourselves that the $|\nabla|$ is normalized to the inverse solar radius. Thus $|\nabla|$ of order unity will imply a structure whose extension is of the order of a solar radius. To make further discussion a little more concrete, let us suppose that we are interested in investigating a structure that has a span $\epsilon R_\odot$, where $\epsilon$ is a number much less than unity. For a structure of order $1000\,$km, $\epsilon\sim 10^{-3}$. The ratio of the orders of various terms in Eq.~(25) are $(|\nabla|\sim L^{-1})$
$$
\begin{array}{ccc}
\frac{\alpha_0^2}{\epsilon^2}\colon &
\frac{\alpha_0}{\epsilon}\,\left(\frac{
1}{a}-d\right)\colon & \left(1-\frac{d}{a}\right) \\
{\rm (1)} & {\rm (2)} & {\rm (3)}\end{array}. \eqno(28)
$$
Of the possible principal balances, the following two are representative:

(a) The last two terms are of the same order, and the first $\ll$ them. Then
$$
\epsilon \sim \alpha_0\,\frac{1/a-d}{1-d/a}.  \eqno(29)
$$
For our desired structure to exist ($\alpha_0\sim10^{-8}$ for $n_0\sim
10^9\,\mbox{cm}^{-3}$), we must have
$$
\frac{1/a-d}{1-d/a}\sim 10^5, \eqno(30)
$$
which is possible if $d/a$ tends to be extremely close to unity. For the first term to be negligible, we would further need 
$$
\frac{\alpha_0}{\epsilon}\ll \frac{1}{a}-d
\Rightarrow\epsilon\gg\frac{10^{-8}}{1/a-d}, \eqno(31)
$$
which is easy to satisfy as long as neither of $a\simeq d$ is close to unity. 
This is, in fact, the standard relaxed state, where the flows are not supposed to play an important part for the basic structure. For extreme sub--Alfv\'enic flows, both $a$ and $d$ are large and very close to one another. Is the new term, then, just as unimportant as it appears to be? The answer is no; the new term, in fact, introduces a qualitatively new phenomenon: Since $\nabla\times (\nabla\times {\bf b})$ is second order in $|\nabla|$, it constitutes a singular perturbation of the system; its effect on the standard root $(2)\sim (3)\gg (1)$ will be small, but it introduces a new root for which the $|\nabla|$ must be large corresponding to a much shorter length scale (large $|\nabla|$). For $a$ and $d$ so chosen to generate a $1000\,$km structure for the normal root, a possible solution would be $d/a\sim 1+10^{-4}$, $d\simeq a=-10$ , then the 
value for $|\nabla|$ for the new root will be (the balance will be from the 
first two terms) 
$$
|\nabla|^{-1}\sim 10^2\,\mbox{cm},
$$
that is, an equilibrium root with variation on the scale of $100\,$cm will be automatically introduced by the flows. The crucial lesson is that even if the flows are relatively weak $(a\simeq d\simeq 10)$, the departure from $\setbox0=\hbox{$\nabla$} \kern-.025em\copy0\kern-\wd0 \kern.05em\copy0\kern-\wd0 \kern-.025em\raise.0433em\box0 \times {\bf B}=\alpha{\bf B}$, brought about by the double $curl$ term can be essential because it introduces a totally different and small scale solution. The small scale solution could be of fundamental importance in understanding the effects of viscosity on the dynamics of these structures; the dissipation of these short scale structures may be the source of primary plasma heating.

We do understand that to properly explain the parallel (to the field--line) motion one must use  kinetic theory since the mean free path along ${\bf B}$ lines can become of the order of $(10^6 - 10^7\,)cm$ for the hot plasma ($100\,eV$).  But since the dissipation acts on the perpendicular energy of the flow, we expect the  two--fluid theory to give  qualitatively (and even quantitatively) correct results.

We would like to remind the reader that by manipulating the force free state $\nabla\times {\bf B}=\alpha({\bf x}){\bf B}$, Parker has built a mechanism for creating discontinuities (short scales) (Parker 1972, 1988, 1994). It is important to note that short length scales are automatically there if plasma flows are properly treated.

(b) The other representative balance arises when we have a complete departure from the one--parameter, conventional relaxed state. In this case, all three terms are of the same order. In the language of the previous section, this balance would demand
$$
\epsilon\sim\alpha_0\ \frac{1}{1/a-d}\sim\alpha_0\
\frac{1/a-d}{1-d/a}   \eqno(32)
$$
which translates as:
$$
\left(\frac{1}{a}-d\right)^2\sim 1-\frac{d}{a}  \eqno(33)
$$
and
$$
\frac{1}{a}-d\sim\alpha_0\ \frac{1}{\epsilon}.  \eqno(34)
$$
For our example of a $1000\,$km structure, $\alpha_0\cdot 1/\epsilon \sim
10^{-5}$, both $a$ and $d$ not only have to be awfully close to one another,
they have to be awfully close to unity. To enact such a scenario, we would
need the flows to be almost perfectly Alfv\'enic. However, let us think of
structures which are on the km or $10\,$km size. In that case $\alpha_0\cdot
1/\epsilon \sim 10^{-2}$ or $10^{-3}$, and then the requirements will become less stringent, although the flows needed are again Alfv\'enic. At a density of $(1 - 4)\cdot 10^8\,\mbox{cm}^{-3}$, and a speed $\sim (200 - 300)\,$km/s, the flow becomes Alfv\'enic for $B_0\sim (1 - 3)\,$G. It is possible that the conditions required for such flows may pertain only in the weak magnetic field regions.

Following are the obvious characteristics of this class of flows:

(1) Alfv\'enic flows are capable of creating entirely new kinds of structures, which are quite different from the ones that we normally deal with. Notice that here we use the term flow to denote not the primary emanations but the plasmas that constitute the existing coronal structures, or the structures in the making.

(2) Though they also have two length scales, these length scales are quite
comparable to one another: This is very different from the extreme sub--Alfv\'enic flows where the spatial length--scales are very disparate.

(3) In the Alfv\'enic flows, the two length scales can become complex
conjugate, i.e., which will give rise to fundamentally different structures in ${\bf b}$ and ${\bf V}$.

Defining $p=(1/a-d)$ and $q=(1-d/a)$, Eq.~(27) can be factorized as 
$$
(\alpha_0\nabla\times -\lambda)(\alpha_0\nabla\times -\mu){\bf
b}=0   \eqno(35)
$$
where $\lambda(\lambda_+)$ and $\mu(\lambda_-)$ are the solutions of the
quadratic equation
$$
\alpha_0\lambda_\pm=-\frac{p}{2}\pm\sqrt{\frac{p^2}{4}-q}. \eqno(36)
$$
If ${\bf G}_\lambda$ is the solution of the equation
$$
\nabla\times {\bf G}(\lambda)=\lambda{\bf G}(\lambda),  \eqno(37)
$$
then it is straightforward to see that
$$
{\bf b}=a_\lambda{\bf G}(\lambda)+a_\mu{\bf G}(\mu),  \eqno(38)
$$
where $a_\lambda$ and $a_\mu$ are constants, is the general solution of the
double $curl$ equation. Using Eqs.~(23'), (37), and (38), we find for the
velocity field
$$
{\bf V}=\frac{\bf b}{a}+\alpha_0\setbox0=\hbox{$\nabla$} 
\kern-.025em\copy0
\kern-\wd0 \kern.05em\copy0\kern-\wd0 \kern-.025em\raise.0433em\box0 
\times
{\bf b}=\left(\frac{1}{a}+\alpha_0\lambda\right)a_\lambda{\bf G}
(\lambda)+\left(\frac{1}{a}+\alpha_0\mu\right)a_\mu{\bf
G}(\mu).  \eqno(39)
$$
Thus a complete solution of the double $curl$ equation is known if we know the solution of Eq.~(37). This equation, also known as the `relaxed--state', or the constant $\lambda$ Beltrami equation, has been thoroughly investigated in literature (in the context of solar astrophysics see for example Parker (1994); Priest (1994)). We shall, however, go ahead and construct a class of solutions for our current interest. The most important issue is to be able to apply boundary conditions in a meaningful manner.

We shall limit ourselves to constructing only two--dimensional solutions. For the Cartesian two--dimensional case ($z$ representing the radial coordinate and $x$ representing the direction tangential to the surface, $\partial/\partial y=0$) we shall deal with sub--Alfv\'enic solutions only. This is being done for two reasons: 1)~The flows in a majority of coronal structures are likely to be
sub--Alfv\'enic, and 2)~this will mark a kind of continuity with the literature. The treatment of Alfv\'enic flows will be left for a future publication.

We recall from earlier discussion that extreme sub--Alfv\'enic flows are characterized by $a\sim d\gg 1$. In this limit, the slow scale $\lambda\sim(d-a)/\alpha_0 \,d\, a$, and the fast scale $\mu=d/\alpha_0$, and the velocity field becomes
\setcounter{equation}{39}
\bq
{\bf V}=\frac{1}{a}\, a_\lambda{\bf G}_\lambda+d a_\mu{\bf
G}(\mu)
\eq
% \eqno(40)
revealing that, while, the slowly varying component of velocity is smaller by a factor $(a^{-1}\simeq d^{-1})$ as compared to the similar part of the magnetic field, the fast varying component is a factor of $d$ larger than the fast varying component of the magnetic field! In a magnetofluid equilibrium, the magnetic field may be rather smooth with a small jittery (in space) component, but the concomitant velocity field ends up having a greatly enhanced jittery component for extreme sub--Alfv\'enic flows (Alfv\'en speed is defined w.r. to the magnitude of the magnetic field, which is primarily smooth, and for consistency we will insure that even the jittery part of the velocity field remains quite sub--Alfv\'enic). We shall come back to elaborate this point after deriving expressions for the magnetic fields.

Equation~(37) can also be written as
\bq
\nabla^2{\bf G}(\lambda)+\lambda^2{\bf G}(\lambda)=0,
\eq
% \eqno(41)
and solving for one component of ${\bf G}(\lambda)$ determines all other
components up to an integration. For the boundary value problem, we will be
interested in explicitly solving for the $z$ (radial) component.

The simplest illustrative problem we solve is the boundary value problem in
which we specify the radial magnetic field $b_z(x, z=0)=f(x)$, and the radial component of the velocity field $V_z(x, z=0)=v_0\, g(x)$, where $v_0\ (\simeq d^{-1}\ll 1)$ is explicitly introduced to show that the flow is quite sub--Alfv\'enic. A formal solution of ($G_z(\lambda)=Q_\lambda$)
\bq
\frac{\partial^2 Q_\lambda}{\partial x^2}+\frac{\partial^2 Q_\lambda}{\partial
z^2}+\frac{\lambda^2}{\alpha^2_0}\, Q_\lambda=0
\eq
% \eqno(42)
may be written as
\bq
Q_\lambda=\int^\infty_{\lambda/\alpha_0} dk\, e^{-\kappa_\lambda z}\ 
C_k\,e^{ikx}+\int_0^{\lambda/\alpha_0} dk\ \cos q_\lambda z\ A_k\, 
e^{ikx}+{\rm c.c.}
\eq
% \eqno(43)
where $\kappa_\lambda=(k^2-\lambda^2/\alpha^2_0)^{1/2}$,
$q_\lambda=(\lambda^2/\alpha^2_0-k^2)^{1/2}$, and $C_k$ and $A_k$ are the expansion coefficients. The equivalent quantities for $Q_\mu$ are $\kappa_\mu$, $q_\mu$, $D_k$, and $E_k$. The boundary conditions at $z=0$ yield (we absorb an overall constant in the magnitude of $b_z$, and $a_\mu/a_\lambda$ is absorbed in $D_k$ and $E_k$):
$$
f(x)=Q_\lambda(z=0)+Q_\mu(z=0),\eqno\mbox{(44a)}
$$
$$
v_0\ g(x)=\frac{1}{a}\ Q_\lambda(z=0)+d\ 
Q_\mu(z=0).\eqno\mbox{(44b)}
$$
Taking Fourier transform (in $x$) of Eq.~(44), we find, after some manipulation, that $(v_0\sim d^{-1},\ |\widetilde f(k)|\simeq|\widetilde g(k)|)$
\setcounter{equation}{44}
\bqy
C_k \simeq \widetilde f(k),  \\
% \eqno(45)
D_k\simeq-\frac{\widetilde f(k)}{d^2}+\frac{v_0}{d}\,\widetilde
g(k)\simeq d^{-2}\widetilde f(k),
\eqy
% \eqno(46)
and functionally (in their own domain of validity) $C_k=A_k$ and $D_k=E_k$.
With the expansion coefficients evaluated in terms of the known functions
(their Fourier transforms, in fact), we have completed the solution for 
$b_z, \ V_z$ and hence of all other field components.

The most remarkable result of this calculation can be arrived at even without a numerical evaluation of the integrals. Although $\widetilde f(k)$ and $\widetilde g(k)$ are functions, we would assume that they are of the
same order $|\widetilde f(k)|=|\widetilde g(k)|$. Then for an extreme
sub--Alfv\'enic flow ($|{\bf V}|\sim d^{-1}\sim 0.1$, for example), the
fastly varying part of $b_z(Q_\mu)$ is negligible ($\sim d^{-2}=0.01$)
compared to the smooth part $(Q_\lambda)$. However, for these very parameters, the ratio 
\bq
\left|\frac{V_z(\mu)}{V_z(\lambda)}\right|\simeq\frac{|C_k/a|}{|d\, 
D_k|}
\simeq\frac{|C_k/a|}{|C_k/d|}\simeq 1;
\eq
% \eqno(47)
the velocity field is equally divided between the slow and the fast scales.
We believe that this realization may prove to be of extreme importance to Coronal physics. Viscous damping of this substantially large as well as
fastly varying flow component may provide the bulk of primary heating needed to create and maintain the bright, visible Corona.

The preceding analysis warns us that neglecting viscous terms in the equation of motion may not be a good approximation until a large part of the kinetic energy has been dissipated. It also appears that the solution of the basic heating problem may have to be sought in the pre--formation rather than the post--formation era. Our time dependent numerical simulation to study the formation of coronal structures was strongly guided by these considerations.

It is evident that for extreme sub--Alfv\'enic flows, the magnetic field, unlike the velocity field, is primarily smooth. But for strong  flows, the magnetic fields  may also develop a substantial fastly varying component. In that case the resistive dissipation can also become a factor to deal with. We shall not deal with this problem in this paper.

Depending upon the choice of $f(x)$ (from which $\widetilde f(k)$ follows)
we can construct loops, arcades and other structures seen in the corona.

\bigskip

\section{\bf Conclusions and discussions}

In this paper we have investigated the conjecture that the structures which comprise the solar corona (for the quiescent Sun) owe their origin to particle (plasma) flows which enter the ``coronal regions'' from lower altitudes. These primary emanations (whose eventual source is likely to be the sun itself) provide, on a continuous basis, much of the required material and energy which constitutes the corona. From a general framework describing a plasma with flows, we have been able to ``derive" several of the essential characteristics of the typical coronal structures.

The principal distinguishing component of the investigated  model is the full treatment accorded to the velocity fields associated with the directed plasma motion. It is the interaction of the fluid and the magnetic aspects of the plasma that ends up creating so much diversity in the solar atmosphere.

This study has led to the following preliminary results:

1. By using different sets of boundary conditions, it is possible to construct various kind of 2D loop and arcade configurations.

2. In the closed magnetic field regions of the solar atmosphere, the primary flows can accumulate, in periods of a few minutes,  sufficient material to build a coronal structure. The ability of the supersonic flows to  generate shocks, and  the viscous dissipation of these shocks can provide an efficient and sufficient source for the primary plasma heating which may take place simultaneously with the accumulation. The stronger the spatial gradients of the flow, the greater is the rate of  dissipation of the kinetic energy into heat. The hot base of the structures is reached at typical distances of a $\sim 10000\,$km from the origin of simulation.

3. A theoretical study of the magnetofluid equilibria reveal that for extreme sub--Alfv\'enic flows (most of the created corona flows) the velocity field can have a substantial, fastly varying (spatially) component even when the magnetic field may be mostly smooth. Viscous damping associated with this fast component could be a major part of the primary heating needed to create and maintain the bright, visible  coronal structure. The far--reaching message of the equilibrium analysis is that neglecting viscous terms in the equation of motion may not be a good approximation until a large part of the  kinetic energy in the primary flow has been dissipated.

4. The qualitative statements on plasma heating, made in points 1 and 2, were tested by a numerical solution of the time--dependent two-fluid system. For sub--Alfv\'enic primary flows we find that the particle-accumulation begins in the strong magnetic field regions (near the solar surface), and soon spans the entire volume of the closed magnetic field region. It is also shown that, along with accumulation, the viscous dissipation of the kinetic energy contained in the primary flows heats up the accumulated material to the observed temperatures, i.e., in the very first (and fast, $\sim (2 - 10)\,min $) stage of accumulation, much of the flow kinetic energy is converted to heat. This happens within a very short distance (transition region) of the solar surface $\sim 0.03R_\odot $. In the transition region, the flow velocity has very steep gradients. Outside the transition layer the dissipation is insignificant, and in a very short time a nearly uniform (with insignificantly decreasing density and temperature on the radial distance), hot and bright quasi-equilibrium coronal structure is created. In this newborn structure, one finds  rather weak flows. One also finds downflows with their parameters determined by the initial and boundary conditions.

The transition region from the solar surface to this equilibrium coronal
structure is also characterized by strongly varying (both radial and across) temperature and density. Depending on the initial magnetic field , the base of the hot region (of the bright part) of a given structure acquires its appropriate density and temperature.

5) The details of the ensuing dynamics are strongly dependent on the relative values of the pressure of the initial flow, and of the ambient solar magnetic field in the region. Two limiting cases were studied with the expected results: 1) The flow entering a relatively weak initial magnetic field strongly deforms (and in specific cases drags) the magnetic field lines, and 2) the flow interacting with a relatively strong magnetic field leaves it virtually unchanged.

\bigskip

We end this paper with several qualifying remarks:

1) This study, in particular the numerical work, is preliminary. We hope to be able to extend the numerical work to make it considerably more quantitative, and to cover a much greater variety of the initial and boundary conditions to simulate the immense coronal diversity. Then a thorough comparison with observations can be undertaken. To show the dissipation of small scale velociy component just like the dissipation of shock--like structures is postponed for future since it requires much higher resolution.

2) This paper is limited to the problem of the origin, the creation and the primary heating of the coronal structures. The processes which may go on in the already existing bright equilibrium corona (secondary or supporting heating, instabilities, reconnection) etc., for example, are not considered. Because of this lack of overlap between our model and the conventional coronal heating models, we do not find it meaningful to compare our work with any in the vast literature on this subject. Led by observations alone, we have constructed and investigated the present model.

3) We do not know much about the primary solar emanations on which this entire study is based. The merit of this study, however, is that as long as they are present (see e.g. [1--3]), the details about their origin are not crucial.

4) We are just beginning to derive the consequences of according a co--primacy 
(with the magnetic field) to the flows in determining overall plasma dynamics. The addition of the velocity fields (even when they are small) brings in essential new physics, and will surely help us greatly in understanding the richness of the plasma behavior found in the solar atmosphere.

\clearpage
{\bf Acknowledgments}

The work was supported in part by the U.S. Dept.~of Energy Contract
No.~DE-FG03-96ER-54346. The study of K.I.N. was supported in part by
Russian Fund of Fundamental Research (RFFR) within a grant 
No.~99--02--18346. The work of N.L.S. was supported in part by the Joint INTAS--Georgia call--97 grant No.~52. S.M.M and N.L.S are also thankful to the Abdus 
Salam International Center for Theoretical Physics at Trieste, Italy.

\newpage

\centerline{\large \bf References}

\begin{enumerate}
\item
%[1]
C.J. Schrijver, A.M. Title, T.E. Berger, L. Fletcher, N.E. Hurlburt, 
R.W. Nightingale, R.A. Shine, T.D. Tarbell, J. Wolfson, L. Golub, J.A.
Bookbinder, E.E. Deluca, R.A. McMullen, H.P. Warren, C.C. Kankelborg, 
B.N. Handy and B. DePontieu, Solar Phys., {\bf 187}, 261 (1999).
\item
%[2]
M.J. Aschwanden, T.D. Tarbell, R.W. Nightingale, C.J. Schrijver, A. 
Title, C.C. Kankelborg, P. Martens and H. P. Warren, Astrophys. J., 
{\bf 535}, 1047 (2000).
\item
%[3]
L. Golub, J. Bookbinder, E. DeLuca, M. Karovska, H. Warren, C.J.
Schrijver, R. Shine, T. Tarbell, A. Title, J. Wolfson, B. Handy and C.
Kankelborg, Phys. Plasmas, {\bf 6}, 2205 (1999).
\item
%[4]
J.M. Beckers,  Ann. Rev. A\&amp;A {\bf 10}, 73  (1972);  Astrophys. 
J., {\bf
203}, 739 (1979).
\item
%[5]
J.D. Bohlin, in {\it Coronal Holes and High Speed Solar Wind Streams},
edited by J.B. Zirker (Colorado Assoc. Univ. Press, Boulder, CO 
1977), 
p.27.
\item
%[6]
G.L. Withbroe, D.T. Jaffe, P.V. Foukal, M.C.E. Huber, R.W. Noyes, E.M.
Reeves, E.J. Schmahl, J.G. Timothy and J.E. Vernazza, Astrophys. J., 
{\bf
203}, 528 (1976).
\item
%[7]
G.L. Withbroe, Astrophys. J., {\bf 267}, 825 (1983).
\item
%[8]
K. Wilhelm, E. Marsch, B.N. Dwivedi, D.M. Hassler, P. Lemaire, A.H.
Gabriel and M.C.E. Huber, Astrophys. J., {\bf 500}, 1023 (1998).
\item
%[9]
G.W. Pneuman and F.Q. Orrall, in {\it Physics of the Sun}, edited by 
P.A. Sturrock (Dordrecht: Reidel, 1986), Vol. II, p.71; K. Shibata, in {\it 
Solar and Astrophysical Magnetohydrodynamic Flows}, edited by K.C.  
Tsinganos (Dordrecht: Kluwer, 1996), p.217; J.H. Thomas, in {\it Solar 
and Astrophysical Magnetohydrodynamic Flows}, edited by K.C. Tsinganos 
(Dordrecht: Kluwer, 1996), p.39; Southwell, K. 1997, Nature, 390, 235
\item
%[10]
H.M. Olussei, A.B.C. Walker II, D.I. Santiago, R.B. Hoover and T.W. 
Barbee Jr., Astrophys. J., {\bf 527}, 992 (1999).
\item
%[11]
U. Feldman, K,G. Widing and H.P. Warren, Astrophys. J., {\bf 522}, 
1133 (1999).
\item
%[12]
H. Peter and P.G. Judge, Astrophys. J., {\bf 522}, 1148 (1999).
\item
%[13]
R. Woo and S.R. Habbal, Astrophys. J., {\bf 510}, L69 (1999).
\item
%[14]
J. D. Scudder, Astrophys. J.,{\bf 398}, 299 (1992).
\item
%[15]
W.C. Feldman, J.T. Gosling, D.J. McComas and J.L. Philips, J. Geophys. 
Res. {\bf 98}, 5593 (1093); W.C. Feldman, S.R. Habbal, G. Hoogeveen and 
Y.-M. Wang, J. Geophys. Res {\bf 102}, 26,905 (1997).
\item
%[16]
X. Li, S.R. Habbal and J.V. Hollweg, J. Geophys. Res {\bf 104}, 2521 
(1999); Y.Q. Yu and S. R. Habbal, J. Geophys. Res. {\bf 104}, 17,045 
(1999).
\item
%[17]
S. Bravo, and G.A. Stewart, Astrophys. J., {\bf 489}, 992 (1997).
\item
%[18]
P.A. Sturrock and R.E. Hartly, Phys. Rev. Lett. {\bf 16}, 628 (1966).
\item
%[19]
M. Banaszkiewicz, A. Czechowski, W.I. Axford, J.F. McKenzie, J.F. and 
G.V. Sukhorukova, 31st ESLAB Symposium. (Noordwijk,   
Netherlands)       ESTEC SP-415, 17, (1997).
\item
%[20]
P. Browning, and E.R. Priest, A\&A, {\bf 131}, 283 (1984).
\item
%[21]
P. Cally, J. Plasma Phys., {\bf 45}, 453 (1991).
\item
%[22]
J.M. Davila, Astrophys. J., {\bf 317}, 514 (1987).
\item
%[23]
J.P. Goedbloed, Phys. Fluids, {\bf 15}, 1090 (1975).
\item
%[24]
M. Goossens, in {\it Advances in Solar System MHD}, 
ed. E.R. Priest and A.W. Hood (Cambridge), 137 (1991).
\item
%[25]
J. Heyvaerts and E.R. Priest, A\&A,{\bf 117}, 220 (1982).
\item
%[26]
J.V. Holweg, Astrophys. J., {\bf 277}, 392 (1984).
\item
%[27]
C. Litwin, and R. Rosner, Astrophys. J., {\bf 499}, 945 (1998).
\item
%[28]
E.N. Parker, Astrophys. J., {\bf 128}, 664 (1998); {\it 
Interplanetary Dynamical Processes} (New York/London: Interscience Publishers 1963); Astrophys. J., {\bf 174}, 499 (1972); Astrophys. J. {\bf 330}, 474 
(1988); J. Geophys. Res., {\bf 97}, 4311 (1992); {\it Spontaneous 
Current Sheets in Magnetic Fields} (Oxford University Press 1994).
\item
%[29]
C.E. Parnell, J. Smith, T. Neukirch, E.R> Priest, Phys. Plasmas, {\bf 
3}, 759 (1996).
\item
%[30]
E.R. Priest and P. Demoulin, J. Geophys. Res., {\bf 100}, 23, 443 
(1995). 
\item
%[31]
E.R. Priest and V.S. Titov, Phil. Trans. R. Soc. Lond.~A., {\bf 354}, 
2951 (1996).
\item
%[32]
I.J.D. Craig and R.B. Fabling, Astrophys. J., {\bf 462}, 969 (1996).
\item
%[33]
E. Galsgaard and A. Nordlund, J. Geophys. Res., {\b 101}, 13445 
(1996).
\item
%[34]
Z. Mikic, D. Schnack and G. Van Hoven, Astrophys. J., {\bf 361}, 690 
(1990).
\item
%[35]
K. Schindler, M. Hesse, and J. Birn, J. Geophys. Res., {\bf 93}, 5547 
(1988).
\item
%[36]
A.A. Van Ballagooijen, Astrophys. J., {\bf 311}, 1001 (1986).
\item
%[37]
J. Heyvaerts and E.R. Priest, A\&A {\bf 137}, 63 (1984); Astrophys. 
J., {\bf 390}, 297 (1993).
\item
%[38]
R.N. Sudan, Phys. Rev. Lett., {\bf 79}, 1277 (1979); R.N. Sudan, 
and D.C. Spicer, Phys. Plasmas, {\bf 4(5)}, 1929 (1997).
\item
%[39]
D. Pfirsch and R.N. Sudan, Phys. Plasmas, {\bf 1(8)}, 2488 (1994); 
{\bf 3(1)}, 29 (1995).
\item
%[40]
S. Tsuneta, Astrophys. J., {\bf 456}, L63 (1996).
\item
%[41]
E.R. Priest, Fifth SOHO Workshop (Oslo) (ESA SP--404), 93 (1997).
\item
%[42]
R. Rosner, W.H. Tucker and G.S. Vaiana, Astrophys. J., {\bf 220}, 643 
(1978).
\item
%[43]
W.M. Neupert, Y. Nakagawa, and D.M. Rust, Solar. Phys., {\bf 43}, 359 
(1975).
\item
%[44]
K.I. Nikol'skaya, Astron. Zh., {\bf 62}, 562 (1985); in {\it Mechanisms 
of Chromospheric and Coronal Heating}, ed. P. Ulmschneider, E. R. Priest 
and R. Rosner (Heidelberg: Springer, 1991), 113.
\item
%[45]
S.R. Habbal, Space Sci. Rev. {\bf 70}, 37 (1994)
\item
%[46]
P. Foukal, {\it Solar Astrophysics} (New York Chichester Brisbane Toronto Singapore: A Wiley--Interscience Publication, 1990).
\item
%[47]
R.D.Richtmyer and K.W.Morton, {\it Difference Methods for Initial--Value
Problems} (Interscience Publishers a division of John Wiley and Sons, 
New York, London, Sydney, 1967).
\item
%[48]
S.T.Zalesak, J.Comp.Phys. {\bf 31}, 335 (1979).
\item
%[49]
S.I.Braginski, {\it Transport Processes in a Plasma}, in Reviews of 
Plasma Physics, edited by M.A. Leontovich (Consultants Bureau, New York, 
1965), Vol.1, p. 205. 
\item
%[50]
J. B. Taylor, Phys. Rev. Lett., {\bf 33}, 1139 (1974); Rev. Mod. Phys., {\bf 58}, 741 (1986).
\item
%[51]
L. Faddeev and Antti J. Niemi, {\it Magnetic Geometry and the 
Confinement of Electrically Conducting Plasma},  Physics/0003083, 
(2000).
\item
%[52]
L. C. Steinhauer and A. Ishida, Phys. Rev. Lett. {\bf 79}, 3423 
(1997).
\item
%[53]
S.M. Mahajan and Z. Yoshida, Phys. Rev. Lett. {\bf 81}, 22 (1998); 
S.M.Mahajan  and Z.. Yoshida, Phys. Plasmas, 7(2), 635-640, 2000; and 
Z.Yoshida, and S.M.Mahajan, Journal Of Mathematical Physics, 40 (10), 5080-5091, (1999).
\item
%[54]
S.M. Mahajan, R. Miklaszewsky, K.I. Nikol'skaya and N.L. Shatashvili, 
{\it Primary Flows, the  Solar Corona and the Solar Wind}. Preprint IFSR 
$\sharp $ 857,Univ.of Texas, Austin, February 1999; "Primary Plasma 
Outflow and the Formation and Heating of the Solar Corona; The High 
Speed Solar Wind Formation". In {\it Structure and Dynamics of the Solar Corona}, eds. B.P.Philipov, V.V. Fomichev, G.N., Kulikova, (Troitsk of Moscow Reg., 2000), p.117.
\item
%[55]
L. Woltjer, in Proc. Nat. Acad. Sci. U.S.A. {\bf }44, 489 (1958).
\item
%[56]
E.R. Priest, ``Magnetohydrodynamics'' in {\it Plasma Astrophysics} by 
J. G. Kirk, D. B. Melrose, E. R. Priest, ed. A.O. Benz and T. J.--L. 
Courvoisier (Springer-Verlag, 1994), p. 1.
\item
%[57]
J.M. Finn and T.M. Antonsen, Phys. Fluids, {\bf 26}, 3540 (1983).

\end{enumerate}

\newpage

\centerline{\large \bf Figure Captions}

\bigskip

Fig.~1

Contour plots for the vector potential $A$ (flux function) in the 
$x-z$ plane for a typical arcade--like solar magnetic field (initial 
distribution). The field has a maximum $B_{max}(x_0=0,\, z_0=0)=7G$\,.

\bigskip

Fig.~2

Hot coronal structure formation by the interaction of the spatially homogenuous
primary flows with 2D arcade--like structure given in Fig.~1\,. The initial
parametrs are: $V_{z0}=300\,km/s$, the temperature and  density of the flow, $T_0=3\,eV$ and $n_0=4\cdot 10^8\,cm^{-3}$ respectively, and the background density $=10^8\,cm^{-3}$. The vector potential $A$, the flow density $n$ 
(normalized to $n_0$), the flow temperature $T$ (in $eV$) and the magnitude
of the flow velocity $|V|$ (in $cm/s$\,) are plotted for $t=750s$ and 
$t=1400s$ . The base of the hot structure is created at a radial distance $\sim 14000\,km$. The distnace scale on the plots is $1=4\cdot 10^{10}cm$\,. The primary heating (and brightening) of the structure is practically stopped in about 23 minutes.

\bigskip

Fig.~3a

The distribution of the radial component $V_z$ (with a maximum of 
$300\,km/s$ at $t=0$\,) for the symmetric, spatially non--uniform velocity field . The plot scale is $1=5\cdot 10^9\,cm$. The process of interaction of such primary flows with the arcade--like magnetic fields (given in Fig.~1 with   
$B_{max}=20\,G$) is accompanied by downflows much slower than the primary flows (plot for $V_z$ at $t=297\,s$). The final parameters of downflows are strongly dependent on the initial and boundary conditions.

\bigskip

Fig.~3b

Hot coronal structure formation by the interaction of the initially symmetric spatially non--uniform primary flows (see plot for $V_z(x,z)$ in Fig.~3a\,) with the 2D arcade--like structure given in Fig.~1\,. Initial parameters are: the temperature and density of the flow, $T_0=3\,eV$ and $n_0=4\cdot 10^8\,cm^{-3}$ respectively, the initial background density $=2\cdot 10^8\,cm^{-3}$, and the field maximum $B_{max}(x_0,\,z_0=0)=20\,G$. The plot scale is $1=5\cdot 10^9\,cm$. The primary heating is completed in a very short time $\sim (2 - 3\,) min$ on distances ($\sim 10000\,km$) shorter than the uniform--flow case when magnetic field was weaker. The heating is symmetric and the resulting hot structure is uniformly heated to $1.6\cdot 10^6\,K$\,. Much of the primary flow kinetic energy has been converted to heat via shock generation.

\bigskip

Fig.~4

The interaction of an initially asymmetric, spatially non-uniform primary flow 
(just the right pulse from the distribution given in Fig.~3a\,) with a strong arcade--like magnetic field ($B_{max}(x_0,\,z_0=0)=20\,G$). Downflows, and the imbalance in primary heating are revealed.

\bigskip

Fig.~5

The interaction of the time--dependent non--uniform initial flow (see plot for the time--distribution of $V_z$ in this Figure; the spatial distribution of the pulse is the same as in Fig.~3a\,) with the arcade--like magnetic field structure (plot in Fig.1 with $B_{max}=20\,G$). The emerging coronal structure has uniform distribution of temperature along the magnetic field (plot for $T$ at $t=371\,s$\,) and the latter is practically undeformed during the formation and heating.

\end{document}